\documentclass[onecolumn,notitlepag,pre,11pt]{revtex4-2}

\usepackage[utf8]{inputenc}
\usepackage{graphicx}
\usepackage{amsmath,amssymb}
\usepackage{hyperref}
\usepackage{xcolor}
\usepackage{bm}

\newcommand{\gbh}{G_\text{bh}}
\newcommand{\Uava}{\langle U \rangle_{a}}
\newcommand{\Uavp}{\langle U \rangle_{p}}
\newcommand{\Dava}{\langle D \rangle_{a}}
\newcommand{\Davp}{\langle D \rangle_{p}}

\DeclareMathOperator{\Tr}{Tr}

\begin{document}
	
	\title{Self-organised structures in mixed active-passive suspensions due to hydrodynamic interactions}
	
	\author{Alexander Chamolly}
	\affiliation{Capital Fund Management S.A., 23 rue de l'Université, F-75007 Paris, France}
	\affiliation{Institut Pasteur, Université Paris Cité, CNRS UMR3738, Developmental and Stem Cell Biology Department, F-75015 Paris, France}
	\affiliation{Laboratoire de Physique de l'École normale supérieure, ENS, Université PSL, CNRS, Sorbonne Université, Université Paris Cité, F-75005 Paris, France}
	
	\author{Takuji Ishikawa}
	\email{t.ishikawa@tohoku.ac.jp}
	\affiliation{Department of Biomedical Engineering, Tohoku University, 6-6-01, Aoba, Aramaki, Aoba-ku, Sendai 980-8579, Japan}
	
	\begin{abstract}
		Microswimmers in suspension exhibit collective swimming behaviour, forming various self-organised structures including ordered, aggregated, and turbulent-like structures. When mixed with passive particles phase-separation is known to occur, but due to the difficulty of accurately handling many-body hydrodynamic interactions, the formation of self-organised structures in mixed suspensions has remained unexplored so far.
		In this study, we investigate the dynamics of mixed dense suspensions of spherical bottom-heavy squirmers and obstacle spheres using Stokesian dynamics in three dimensions, taking hydrodynamic interactions into account. The results show that without an external orientating mechanism the formation of orientational order is in general disturbed by the presence of passive spheres. An initially phase-separated state is metastable for neutral or puller squirmers at high packing densities. When the squirmers are bottom-heavy, phase-separation can occur dynamically in some cases, notably a fibrillar kind of separation for neutral squirmers and pullers at medium densities. We also observed a novel form of lamellar phase-separation for pullers at high densities with strong bottom-heaviness, with a sandwich-like structure consisting of a layer of passive particles pushed by a layer of swimmers, followed by a gap.
		These results indicate that microstructure and particle transport undergo significant changes depending on the type of swimmer, highlighting the importance of hydrodynamic interactions.
		These insights allow for a deeper understanding of the behaviour of active particles in complex fluids and to control them using external torques.
	\end{abstract}
	
	
	\maketitle

	\section{Introduction}

	Microswimmers exhibit diverse collective behaviours when suspended in liquids, such as bioconvection \citep{Pedley1992,Hill:2005aa,Bees:2020aa}, bacterial turbulence \citep{Aranson_2022,Alert2022}, and migration due to taxis \citep{Ishikawa2025taxis}.
	The structure of these collective behaviours has been analysed extensively using both continuum and discrete models of microswimmers.
	Among these, discrete models have the advantage of being able to describe phenomena at a high resolution because they can express the movement of each individual.
	Two major examples are active Brownian particles (ABPs) and squirmers \citep{Moran2017, ishikawa2025perspectives}.
	The ABP model is one of the simplest mathematical frameworks with which to describe the behaviour of active colloids. In its most basic form, the model disregards hydrodynamic and phoretic interactions, and considers only steric repulsion between particles \citep{Stark2023}. An ABP moves at a constant speed in the direction of its orientation, which fluctuates due to rotational Brownian motion. Its position and orientation evolve under the influence of both translational and rotational diffusion, driven by thermal noise.
	From a computational perspective, the evolution of an ensemble of ABPs can be parallelised with an explicit time step, making ABPs favourable for studying large ensembles.
	Despite these simplifications, ABPs can exhibit rich collective dynamics, including motility-induced phase-separation (MIPS), clustering and turbulent-like patterns \citep{Bialke2012,Bialke2013,Fily2012,bechinger2016active,Stark2023}.
	Large-scale analyses using a coarse-grained model of rod-like swimmers have also been conducted, with reports on orientation instability and the formation of turbulent-like structures \citep{Saintillan2007,Krishnamurthy2015,Bardfalvy2019}.
	
	In contrast, the \textit{squirmer} model prescribes a surface velocity on the microswimmers which is interpreted as a boundary condition for the interstitial fluid flow. \citep{Lighthill:1952, Blake:squirmer, Pedley2016IMA,Ishikawa2024ARFM}.
	As such, this model can capture the short- and long-range hydrodynamic interactions (HIs) inherent in the relevant Stokes flow regime.
	This modelling approach involves solving coupled equations for many particles simultaneously, which is computationally expensive, but provides a more accurate description of swimmer interactions.
	In a comparative study, \cite{Theers2018} simulated both ABPs and squirmers in quasi-two-dimensional (2D) geometry. Under the same conditions, spherical ABPs formed larger and denser clusters than squirmers. The suppression of MIPS in squirmers is attributed to re-orienting torques induced by HIs during collisions. Furthermore, the degree of MIPS suppression varies with swimmer type, being strongest in pushers and weaker in pullers.
	The tendency for hydrodynamic interactions to suppress MIPS has also been reported in two-dimensional disc squirmers \citep{Matas2014} and three-dimensional dense suspensions of squirmers \citep{zhou2025}.
	These results demonstrate the significant impact of HIs on the microstructure of active suspensions.

	Due to the challenges of modelling and interpreting dynamics in full three dimensions (3D), the hydrodynamic emergence of collective dynamics has been studied most extensively in quasi-2D suspensions. Here \cite{ishikawa2008coherent} revealed the existence of dynamic clustering and mesoscale spatiotemporal patterns of squirmers. Furthermore, when a gravitational torque (bottom-heaviness) is introduced, the squirmers can align and form horizontal bands.
	The rheological properties of thin-layered squirmer suspensions \citep{Pagonabarraga2013}, as well as their orientation and cluster formation \citep{Alarcon2019}, have also been reported.
	\cite{Kyoya2015} investigating the effect of swimmer shape. They found that neutral squirmers tend to exhibit polar order, which is disrupted as the swimming mode becomes more pusher- or puller-like. An increased aspect ratio suppresses the polar order, but enhances mesoscale chaotic behaviour. Near-field hydrodynamic interactions were identified as the primary mechanism underlying these behaviours.
	Further work by \cite{Zantop2022} studied rod-like squirmers in narrow slits. These swimmers exhibited various collective behaviours, including turbulence, dynamic clusters and swarms. The resulting state depended strongly on the aspect ratio and volume fraction of the swimmers. Higher aspect ratios and densities resulted in more pronounced clustering and the emergence of coherent structures. 
	
	Fully 3D suspensions provide a more complete picture of hydrodynamically mediated collective behaviour but have been more rare. \cite{ishikawa2008development} used Stokesian dynamics to simulate spherical squirmers in unbounded suspensions. Their results revealed the spontaneous emergence of polar order. This ordering was strongest in neutral squirmers and weakest in pushers.
	Subsequent work by \cite{Yoshinaga2018} showed that this polar order mainly arises from short-range lubrication interactions rather than long-range flows. The ordering remains stable even in large systems with finite noise. \cite{Oyama2016} observed a similar alignment of squirmers when they were confined between parallel plates. Interestingly, these squirmers aligned perpendicular to the walls, forming coherent travelling clusters that reflected off the boundaries and propagated like waves.
	In an even more complex scenario, \cite{Samatas2023} studied helical squirmers, which exhibit both translational and rotational motion. Starting from random configurations, these swimmers spontaneously synchronised their spinning and aligned their rotational axes due to hydrodynamic coupling, forming collective helical swimming states.
	
	Recent research has revealed the behaviour of microswimmers, such as microalgae and bacteria, in more complex environments, such as porous media \citep{Brun2019,dentz2022dispersion,Dehkharghani2023}, polymer solutions \citep{zottl2019enhanced}, and mixtures of cells with different properties \citep{meacock2021bacteria}.
	This knowledge is crucial to understand how microorganisms behave in complex environments where they actually live, such as the intestines and soil.
	In anticipation of such complex settings, mixed suspensions of active and passive particles have also been investigated \citep{bechinger2016active,wang2019interactions}. Here, the effect of few active particles in an otherwise inert phase is called `active doping' and can affect crystallisation behaviour. Conversely, a few passive tracers in an `active bath' exhibit interesting diffusive and clustering behaviour. For dense mixtures, both MIPS and turbulent behaviour have been observed. However, many analytical studies have focused on ABPs that do not take hydrodynamic interactions into account. Examples include 2D clustering due to the active doping effect \citep{van2016fabricating,wittkowski2017nonequilibrium,alaimo2018microscopic}, 2D MIPS \citep{stenhammar2015activity,wysocki2016propagating,dolai2018phase,ai2018mixing,ilker2020phase,kolb2020active,Rogel2020}, and diffusion \citep{,kim2022active}.
	
	On the other hand, hydrodynamic interactions have been present in relevant experimental studies. In quasi-2D settings, \cite{Kummel2015} demonstrated that passive particles form clusters when a small number of self-diffusiophoretic microswimmers with a diameter of 4.2 $\mu$m are introduced into a dense suspension of passive particles with a similar diameter. Even at an active particle volume fraction of just 0.01\%, collisions with passive particles generate space around the swimmers, driving the passive particles to aggregate into clusters.
	\cite{Gokhale2022} observed similar clustering in bacterial suspensions of \textit{Pseudomonas aurantiaca} and \textit{Escherichia coli} using 3.2 $\mu$m colloids. Without bacteria, the colloids were evenly distributed; however, significant clustering emerged as the bacterial concentration increased, even below the threshold for active turbulence. Unlike tracer diffusion, which is caused by swimmer-generated flow fields, clustering arises from size-dependent interactions between swimmers and tracers. Finally, quasi-2D mixed suspensions of active and passive particles have also been used to investigate diffusion \citep{aragones2018diffusion,Wu2000,Peng2016,alonso2019transport}, and collective motions \citep{pessot2018binary}.
	
	In 3D settings with HIs, \cite{chamolly2017active} systematically analysed the behaviour of a squirmer in a lattice of rigid spheres, mapping swimmer behaviour in the parameter space defined by the squirmer type and the lattice packing density. Puller-type squirmers always travelled in near-straight lines, regardless of density. Weak pushers also moved in a straight line in low-density lattices, but exhibited random movement at higher densities due to increased interactions. When pushers became strong, they transitioned to a trapped state, either orbiting a single obstacle or becoming immobilised. This study demonstrated how lattice structure and swimmer type determine the behaviours of microswimmers.
	Recently, \cite{ge2025hydrodynamic} investigated the hydrodynamic diffusion of mixed suspensions of non-self-propelling spherical squirmers and rigid spheres using Stokesian dynamics \citep{Elfring_Brady_2022}. They found that apolar active suspensions are most diffusive when the volume fraction is between 0.1 and 0.2.
	3D diffusion of tracer particles in microswimmer suspensions have also been reported \citep{Delmotte2018,Kogure2023}.
	However, the formation of self-organised structures through hydrodynamic interactions in 3D mixed suspensions has so far remained unclear due to the difficulty of accurately handling many-body HIs.
	
	Therefore, in this study, we investigate 3D mixed suspensions of spherical squirmers and obstacle spheres, taking hydrodynamic interactions into account.
	The near- and far-field hydrodynamics, including lubrication forces, are accurately analysed by using Stokesian dynamics. The methodology is explained in detail in \S \ref{method}.
	In \S \ref{sec:results}, we investigate the orientational order, phase-separation and microstructures in 3D mixed suspensions by varying the swimmer types, packing densities, proportion of active and passive particles, bottom-heaviness, and initial conditions.
	Finally in \S \ref{discuss}, we compare the microstructures obtained in this study with those obtained in previous studies.

	\section{Methods}
	\label{method}
	
	\begin{figure}[th!]
		\centering
		\includegraphics[width=\columnwidth]{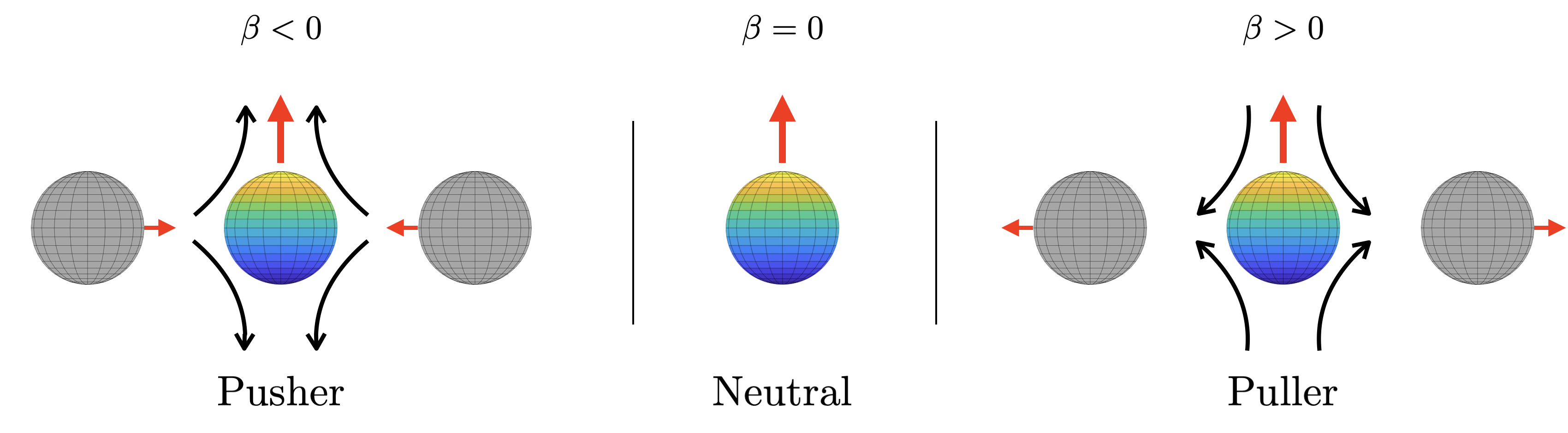}
		\caption{Sketch of the long-range interactions between an upward moving active swimmer and passive spheres, as a function of the squirmer parameter $\beta$. At leading order, a pusher attracts passive objects laterally, while a puller repels them. In contrast, short-range interactions are governed by lubrication flows for all swimmer types.}
		\label{fig:illust}
	\end{figure}
	
	\subsection{Numerical framework}
	The microswimmers and inert particles are all modelled as identical non-Brownian spheres with radius $a$ scaled to unity. Active spheres feature a surface velocity of the form
	\begin{align}
		u_r(1,\theta) =u_\phi(1,\theta) = 0, \quad u_\theta(1,\theta) = \frac{3}{2}U_0\sin\theta\left(1+\beta\cos\theta\right),
	\end{align}	
	in the swimmer body frame, and exert a force-dipole in the far-field, the strength and sign of which is determined by the squirmer parameter $\beta$ (Figure~\ref{fig:illust}). The velocity magnitude $U_0$ of a solitary microswimmer is assumed to be small, resulting in swimming at a very small Reynolds number.
	From here on, all physical quantities are nondimensionalised by using the characteristic length $a$, the characteristic velocity $U_0$, and the viscosity $\mu$.
	
	Optionally, we consider that swimmers may be bottom-heavy with a restoring torque $\bm{G}_{\text{bh}}=\tfrac{4}{3} \pi \rho a^3 h\bm{e}\times\bm{g}$ acting on the swimmer, where $h$ the distance of the centre of mass from the geometrical centre along the orientation of the swimmer $\bm{e}$, $\rho$ is the swimmer density and $\bm{g}$ the gravitational vector. In this paper, we take the maximal dimensionless scalar value of the torque, $G_\text{bh} = \tfrac{4}{3} \pi \rho a h |\bm{g}| / \mu U_0$, as the bottom-heaviness parameter.
	
	Passive spheres instead feature a no-slip boundary condition and do not self-propel. All spheres are neutrally buoyant, force- and torque-free, and their instantaneous velocity and rotation vectors are determined from the solution of the resulting linear system of equations.
	
	To accurately analyse the hydrodynamic interactions of an infinite number of particles in Stokes flow, we use a robust computational framework that has been described in detail in previous work \citep{ishikawa2012vertical}. Briefly, we employ Stokesian dynamics \citep{banchio2003accelerated,ishikawa2008development} that simultaneously accounts for far- and near-field HIs. An infinite number of reflected far-field interactions between a periodically infinite number of microswimmers are calculated by Ewald summation technique \citep{beenakker1986ewald}, while near-field interactions between particles are accounted for using lubrication theory and a pre-computed database of lubrication interactions \citep{ishikawa2006hydrodynamic}. A short-range repulsive force models steric interactions to prevent particles from overlapping. This allows us to accurately model hydrodynamic interactions at both short and long separation distances in 3D. A sketch of the mathematical underpinnings, with references, is presented in \S\ref{appendix:numericalmethod}.

	\subsection{Choice of parameters and boundary conditions}
	We simulate a suspension of $N=216$ particles in a 3D-periodic domain. The particles are initialised on a hexagonal close-packed lattice of $6 \times 6 \times 6$ spheres, and the volume fraction of objects $\phi$ is set by the lattice spacing. In this way we strike a balance between computational feasibility, and a sufficiently large number of interactions and size of the computational domain to suppress unwanted finite-size effects.
	
	Active swimmers are initialised with pseudo-random initial orientation. For an initially mixed suspension, a fraction $\alpha$ of objects is chosen as active, and chosen at random from all $N$ spheres.
	For an initially phase-separated suspension, starting from the original $6 \times 6 \times 6$ lattice, $6\alpha$ consecutive lattice planes are initialised with active particles, followed by $6(1-\alpha)$ planes initialised with passive particles.
	
	We choose the following range of values for the parameters: For the active fraction we consider $\alpha=\{1/6,1/2,5/6,1\}$. We omit $\alpha=0$ because there is no motion in a neutrally buoyant passive non-Brownian suspension, and we note that $\alpha=1$ has been described previously in studies on purely active suspensions \citep{ishikawa2008development}, so it is included here only for comparison. For the squirmer parameter we consider $\beta=\{-3,-1,0,1\}$. This asymmetric choice is motivated by previous work \cite{chamolly2017active} that identified a qualitative difference between strong and weak pushers in their interaction with stationary obstacles, but no equivalent behaviour for pullers. For the packing density, i.e.\ volume fraction, we consider $\phi=\{0, 0.1, 0.2, 0.3, 0.4, 0.5\}$, with the upper limit imposed by numerical constraints. The maximum packing density of equal spheres is $\phi_\text{max}=\pi/3\sqrt{2}\approx 0.74$, which would prohibit any sort of rearrangements. In practice we find that at $\phi=0.5$ rearrangements are still possible but steric interactions play an important role in the dynamics. For the bottom-heaviness parameter we choose $\gbh=\{0, 10, 100\}$, corresponding to no, weakly, and strongly preferred orientation of the active swimmers.	
	
	We simulate the system for 200 time units with a time step $dt$ of 0.001 units, except for $\phi=0.1$ with $\gbh=0$ where we simulate 1000 time units to definitively establish the existence of an ordered state.
	The time stepping is performed with the fourth-order Adams-Bashforth scheme. Unless otherwise stated in the discussion of results, we find that at this point the system has reached a steady state, with the relevant quantities of interest varying only subject to the randomness of the initial conditions.
	
	\subsection{Evaluation metrics}\label{sec:metrics}
	
	To facilitate an easy read of the results section \ref{sec:results}, we summarise and define here the quantities of interest that we can calculate and that characterise the behaviour of the suspension. In particular, we focus on the coherence of emerging structures, as well as phase-separation.
	
	In general, we denote by $\langle .\rangle_a$ and $\langle .\rangle_p$ an average over all active and passive particles, and let $N_a$ and $N_p$ indicate their respective counts ($N_a + N_p = N = 216$). Active swimmers have an orientation vector $\bm{e}$ that indicates the direction of their propulsion. The mean orientation vector $\langle \bm{e} \rangle_a(t)$ is then a meaningful measure of coherence, with $|\langle \bm{e} \rangle_a|\approx 1$ indicating strong alignment. We can similarly define the mean velocities $\langle \bm{U} \rangle_{a,p}(t) := \delta t^{-1}\langle \bm{r}(t+\delta t) - \bm{r}(t) \rangle_{a,p}$ with $\delta t = 0.1$, which also apply more generally to passive particles that are advected through hydrodynamic interactions. In our analysis we found that $\langle \bm{e} \rangle_a$ and $\langle \bm{U} \rangle_a$ are highly correlated,
	which indicates an absence of jamming. For sake of using a comparable measure for both kinds of particle, we will therefore focus on mean velocity and in particular, with a slight abuse of notation, the scalar quantities $\langle U \rangle_{a,p} = |\langle \bm{U} \rangle_{a,p}|$. Additionally, in the presence of bottom-heaviness, it is interesting to consider the cross-sectional number flux of passive particles in the direction of the external director field, which is given by $Q_p= N_p \langle - \hat{\bm{g}} \cdot \bm{U} \rangle_p \times\phi/ \tfrac{4N}{3}\pi a^3$.
	
	While non-zero mean velocities are expected to emerge for coherent motion and transport, chaotic dynamics in which particle motion is more akin to a random walk are better characterised by a measure of diffusion. To this end, we define the (translational) dispersion tensor,
	\begin{align}
		\langle \bm{D} \rangle_{a,p}(t) := \frac{\langle \left[\bm{r}(t) - \bm{r}(0)\right] \left[\bm{r}(t) - \bm{r}(0)\right]\rangle_{a,p}}{2 t}.
	\end{align}
	As outlined in previous work on homogenous suspensions \citep{ishikawa2007diffusion}, the scaling of the diffusion tensor with $t$ as $t\to\infty$ informs us whether the suspension behaves diffusively or ballistically, in which case $\langle \bm{D} \rangle_{a,p}\to \text{const.}$ or $\sim t$, respectively. Moreover, the trace $\langle D \rangle_{a,p} = \tfrac{1}{3} \langle \Tr\bm{D} \rangle_{a,p} = \tfrac{1}{6t} \text{MSD}_{a,p}$ is directly related to the more intuitive mean-squared displacement (MSD). However, we note that in the case of bottom-heaviness and coherent order $\langle \bm{D} \rangle_{a,p}$ is in general not isotropic. To confirm this, we computed the anisotropy of the dispersion tensor as $A = \sqrt{\tfrac{3}{2} \left(\sum_{i,j}D_{ij}^2 / (\Tr\bm{D})^2 - \frac{1}{3}\right)}$, and found that it generically increases from near 0 (isotropic up to noise) to near 1 (essentially one-directional motion) in all cases where orientational order emerges. Since this does not usefully characterise the nature of the ordered state between the two phases in more detail, we instead developed another measure in terms of spatial correlations, defined below.
	
	We note that in the calculation of the displacement vectors for both $\langle U \rangle_{a,p}$ and $\langle D \rangle_{a,p}$, it is necessary to take into account the periodicity of the simulated domain to avoid discontinuities introduced by particles crossing the boundaries of the simulation. Likewise, for the calculation of the instantaneous pairwise separation, the periodicity is equally taken in account by also calculating the distance to all surrounding periodic copies and selecting the smallest of these values.
	
	Finally, we would like to be able to quantify whether there are any interesting structures forming in the mixed suspension, and in particular, whether phase-separation is dynamically achieved or at least maintained. By definition, phase-separation is a phenomenon of statistical physics that is best defined in the limit where the number of particles is infinite, or can be at least approximated by a continuous phase field. Here, we are constrained numerically to a rather small number of 216 particles in 3D, which makes it challenging to define a volume density of active and passive spheres. For this reason we proceed as follows. First, we choose one spatial direction to project into, and so reduce the dimensionality of the problem. Then, we bin active and passive particles in a grid $(i,j)$ of $m_i \times m_j$ subdivisions of the computational domain. Let $n_a(i,j)$ and $n_p(i,j)$ denote the number of particles counted in each bin. Using the mean $\bar{n}_*$ and standard deviation $\sigma_{n_*}$ of each particle type, we define normalised counts
	\begin{align}
		z_a(i,j) = \frac{n_a(i,j) - \bar{n}_a}{\sigma_{n_a}} , \quad z_p(i,j) = \frac{n_p(i,j) - \bar{n}_p}{\sigma_{n_p}}
	\end{align}
	and introduce the adjacency kernel
	\begin{align}
		W(i,j;k,l) = \begin{cases}
			1/8 & \text{if}\ |i-k| \leq 1\ \text{and}\ |j-l| \leq 1\ \text{and not}\ \left( i=k\ \text{and}\ j=l\right),\\
			0 & \text{otherwise}
		\end{cases},
	\end{align}
	which, when convolved with a density, replaces the value in each cell with the average over all 8 neighbouring cells. This allows us to define the spatial cross-correlation $S$ as
	\begin{align}
		S = \frac{1}{m_im_j}\sum_{i,j}\sum_{k,l} z_a(i,j)W(i,j;k,l)z_p(k,l).
	\end{align}
	As shown in appendix \ref{appendix:sproperties}, $S^2\leq1$. In the case of phase-separation, we expect the fields $z_a$ and $z_p$ to be anti-correlated, and thus $S$ to be negative, with more strongly negative values for more clearly defined separation.
	
	We emphasise that this metric relies on subjective choices, in particular the choice of direction to project in, the grid resolution, and the convolution kernel. Regarding the direction, it turns out that there is usually a clear direction of phase-separation in our simulations, either by virtue of the initial condition, or the direction of gravity, hence this choice does not pose a problem in this context. Experimentation with the grid resolution showed that $m_i=m_j=12$ yields a good trade-off between resolution of the grid and particle counts (with a mean count of $1.5$ spheres per bin). The kernel $W$ is chosen to account for rotational symmetry and detect phase-separation patterns on the length scale of the grid resolution, which is on the order of a sphere radius. In practice, we have found that these choices enable $S$ to capture quite well the subjective impression from plotting the suspension both in 3D and 2D.
	
	\section{Results}\label{sec:results}
	
	In total, the four parameters $\{\alpha, \beta, \phi, \gbh\}$, together with two different initial configurations (mixed and phase-separated) give a five-dimensional space to explore and analyse based on several different quantities of interest. Due to the correspondingly large number of parameter combinations, we do not document each quantity defined in \S\ref{sec:metrics} exhaustively, but instead focus on the most interesting features in each region of parameter space and show figures with illustrations and the most appropriate quantities of interest. Additionally in the officially published version of this article  \cite{chamolly2026self}, we provide animations of the suspensions in the form of supplementary videos. These are able to convey the formation of the different structures much better than is possible with only text and static images. Their captions may be found in \S\ref{appendix:moviecaptions}. Please note due to an error in the editorial process the videos in the published version are in the order 1, 10, 2-9.
	
	\subsection{No bottom-heaviness ($\gbh=0$)}	
	\subsubsection{Initially mixed suspensions: coherent order and diffusion}
	
	\begin{figure}[th!]
		\centering
		\includegraphics[width=\columnwidth]{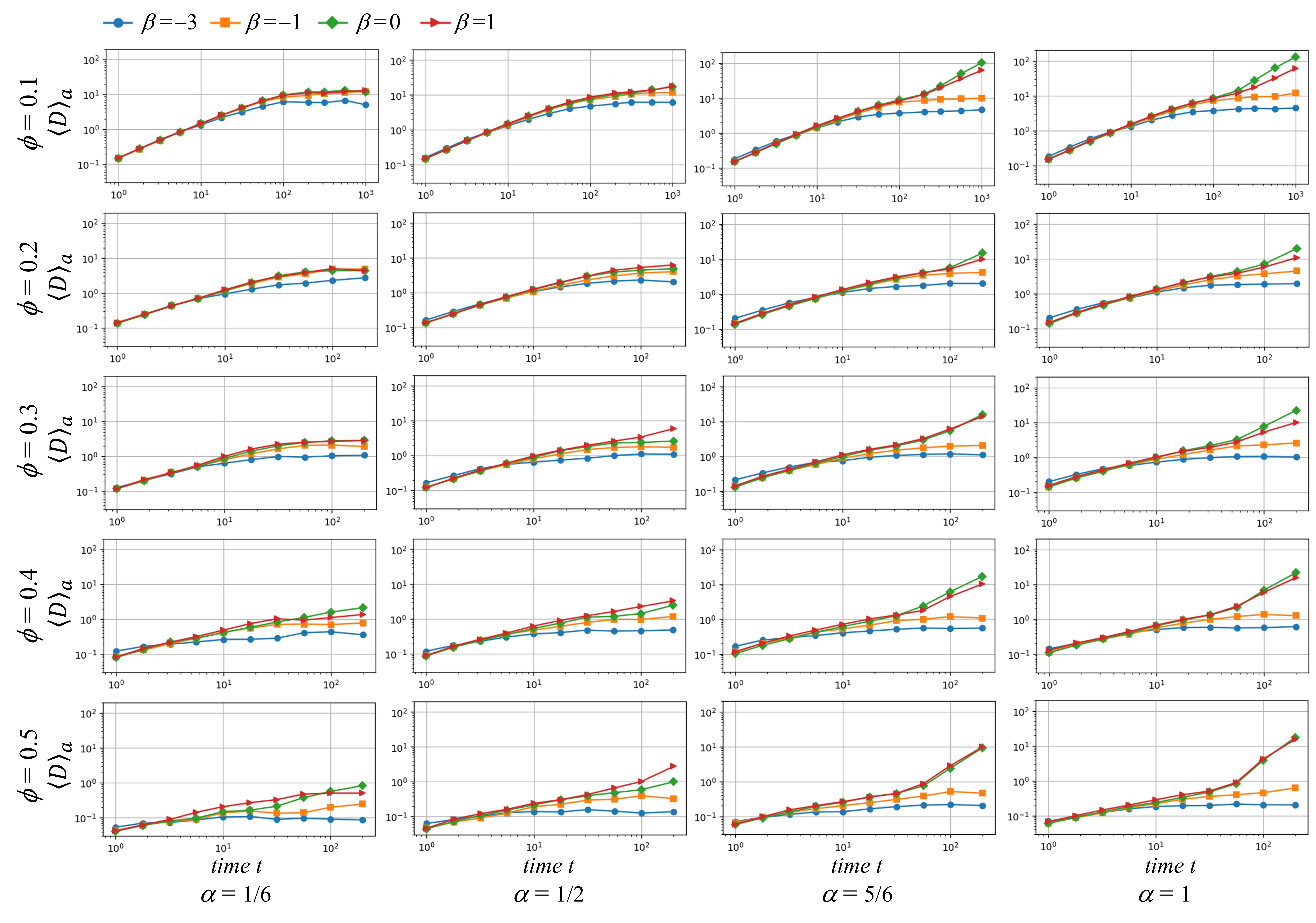}
		\caption{Temporal evolution of the dispersion coefficient of active particles $\Dava$ of four types of squirmers from an initially mixed state with random orientations.
			The active fraction $\alpha$ is varied from $1/6$ to 1, and the volume fraction $\phi$ is varied from 0.1 to 0.5. For $\phi=0.1$, the computation time was extended to $t=1000$.}
		\label{fig:Davaevo}
	\end{figure}
	
	\begin{figure}[th!]
		\centering
		\includegraphics[width=\textwidth]{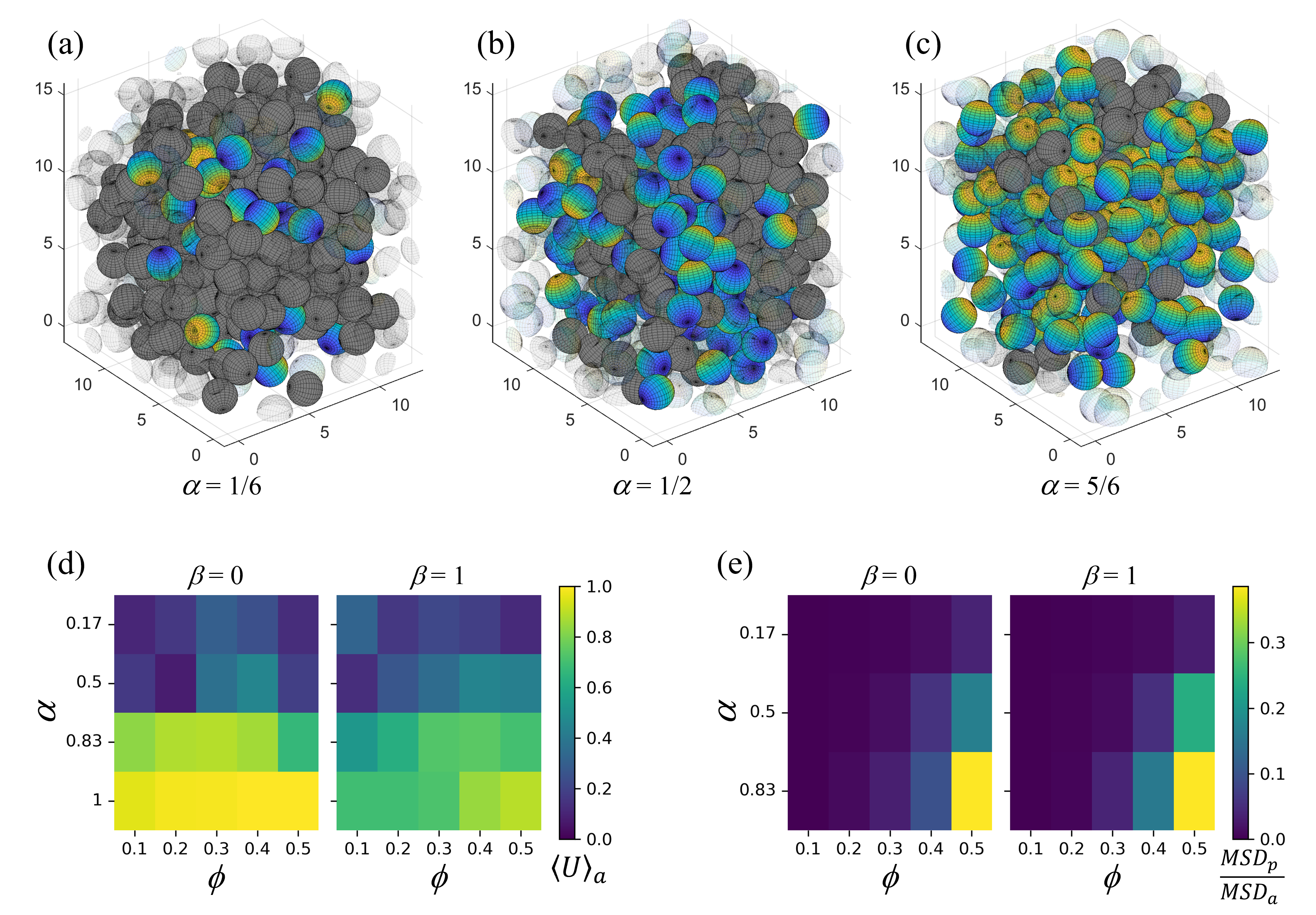}
		\caption{Coherent structures in initially mixed suspensions.
			(a-c) Steady state at $\phi=0.40$, $\beta=1$ for different values of active fraction $\alpha$: (a) $\alpha=1/6$, (b) $\alpha=1/2$, (c) $\alpha=5/6$. See also supplementary movies 1-3.
			(d) Mean active velocity $\Uava$ at $t=200$.
			The presence of passive objects interferes with the formation of a coherent structure, which is only present for $\alpha\geq5/6$. 
			(e) Ratio of mean-squared displacements $\text{MSD}_p / \text{MSD}_a$ at $t=200$.
			The MSD of passive particles $\text{MSD}_p$ is a small fraction of that of the active swimmers $\text{MSD}_a$.}
		\label{fig3}
	\end{figure}
	
	It is well established that without the inclusion of bottom-heaviness or another mechanism to break isotropy no orientational order emerges in a suspension of pushers ($\beta< 0$), and moreover that any initially ordered state is unstable \citep{ishikawa2008coherent,ishikawa2008development}. As such, it is not very surprising that the same holds true for pushers in the presence of passive particles. We find that orientational correlations are approximately zero, and squirmers
	move diffusively, with $\Dava$ reaching a plateau. Moreover, this active diffusivity decreases with $\alpha$, $|\beta|$ and $\phi$ as more frequent near-field interactions between squirmers increase the number of reorientation events (Figure~\ref{fig:Davaevo}). 
	
	In contrast, in both pure neutral and pure puller suspensions ($\alpha = 0, \beta\geq0$), a state of collective motion emerges \citep{ishikawa2008development} for all packing densities in the parameter range (Figure~\ref{fig:Davaevo}). 
	Under these conditions, squirmers show a coherent order, and $\Dava$ does not converge but increases over time, exhibiting ballistic behaviour.
	Here we find that the presence of passive objects interferes with the formation of the coherent order (Figure~\ref{fig3}a-c and supplementary movies 1-3). For active fraction $\alpha\leq 0.5$, the average mean velocity $\Uava < 0.5$ in the steady state, with slightly higher values for pullers than for neutral swimmers (Figure~\ref{fig3}d). At $\alpha = 5/6$, on the other hand, a coherent state does form for both types of swimmers regardless of volume fraction $\phi$. Moreover, the steady state coherence of this state, as measured by $\Uava$, is only slightly reduced from that of a pure squirmer suspension with $\alpha=1$ except at very high volume fraction ($\phi=0.5$) where steric interactions between passive obstacles and active squirmers play a role.
	
	It may be expected that regardless of whether coherent structures form, the flows created by the squirmers lead to a mixing of the passive part of the suspension. In practice, this effect is extremely weak in the sense that $\text{MSD}_p / \text{MSD}_a \lesssim 5\%$ at $t=200$ (Figure~\ref{fig3}e). When no coherent structure forms, the motion of the passive particles is indeed diffusive, with $\Uavp$ reaching a plateau similarly to $\Uava$. In the presence of an ordered state however, there is a dependence on the packing density $\phi$. While passive obstacles in a dilute suspension ($\phi\leq 0.3$) continue to move diffusively, at high volume fractions they adopt the ballistic behaviour of the coherent active phase, with $\Davp$ growing linearly in time and $\Uavp$ eventually becoming comparable to $\Uava$ at $\phi=0.5$. This effect is a little stronger for pullers than for neutral squirmers, presumably due to the lateral attraction they induce (Figure~\ref{fig:illust}), but its emergence at high volume fractions suggests that it is due to a dominant role of steric interactions and limited space for squirmers to move around passive obstacles. 
	
	\subsubsection{Initially phase-separated suspensions: the stability of phase-separation}
	\begin{figure}[th!]
		\centering
		\includegraphics[width=\textwidth]{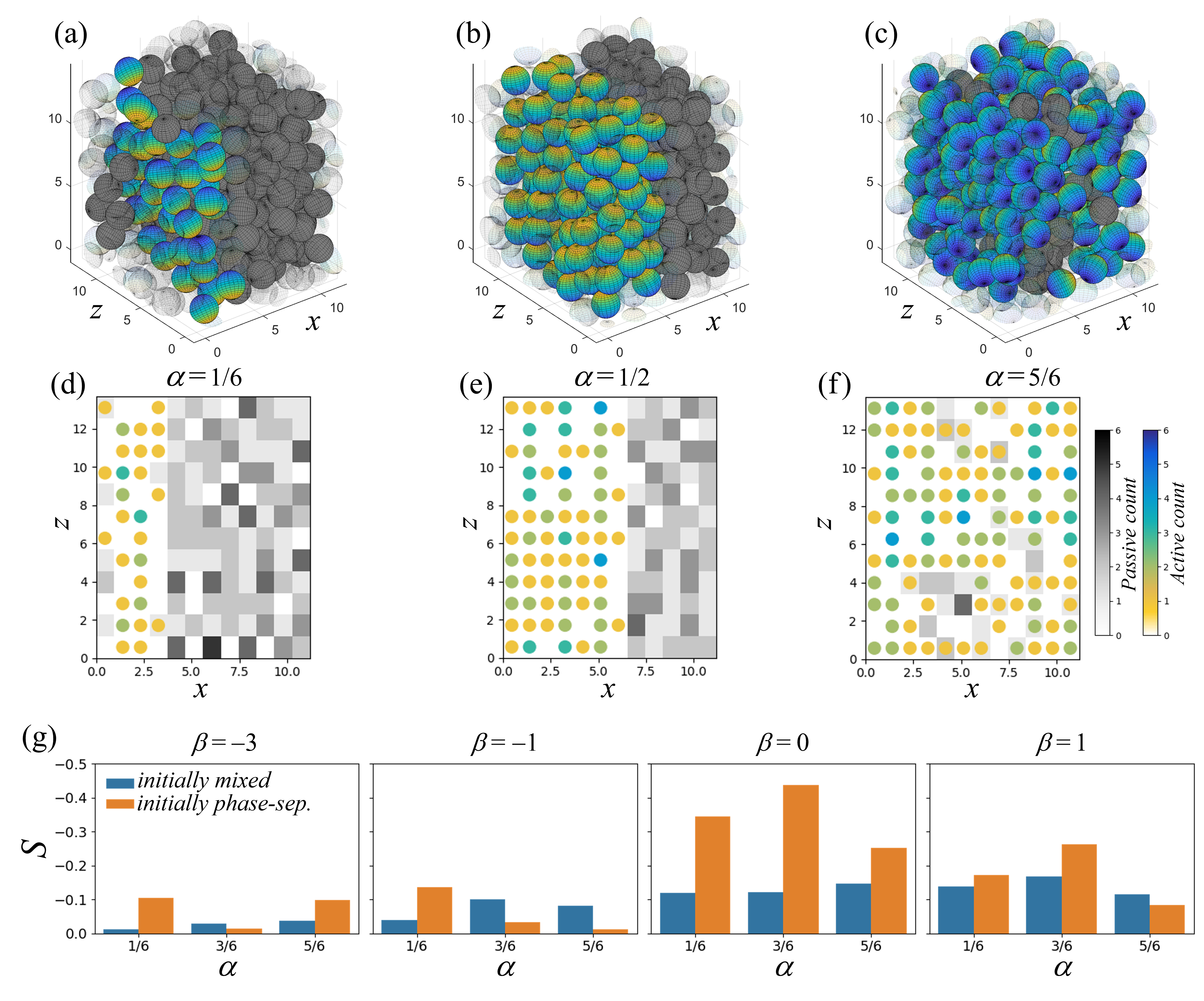}
		\caption{Stability of phase-separation ($\phi=0.5$).
			(a-c) Illustration of the steady state $(t\approx 180)$ for initially phase-separated suspensions of neutral squirmers ($\beta=0$) with (a) $\alpha = 1/6$, (b) $\alpha = 1/2$, and (c) $\alpha = 5/6$. See also supplementary movies 4-6. For $\alpha \leq 1/2$ the passive spheres act as a quasi-2D confinement to a layer of neutral squirmers in which a coherent structure forms. For $\alpha=5/6$, the passive layer is destroyed.
			(d-f) Density projection into a plane orthogonal to initial phase-separation, and sphere counts in a 2D grid of bins that form the basis of the computation of spatial cross-correlation $S$ with (d) $\alpha = 1/6$, (e) $\alpha = 1/2$, and (f) $\alpha = 5/6$ ($\beta=0$).
			(g) Steady-state 2D spatial cross-correlation $S$ for initially mixed vs initially phase-separated suspensions at $t=180$. Large negative values indicate phase-separation between active and passive spheres.}
		\label{fig4}
	\end{figure}
	
	As described so far, we did not observe a separation of active and passive phases in any of our simulations of an initially mixed suspension without bottom-heaviness. To confirm whether such a state could still exist or would be unstable, we performed the same simulations again with an initial partition of the simulation domain into a purely active and a purely passive phase. With this setup we found that for most of the parameter space the separation is indeed unstable. At low volume fractions $\phi$, active particles move on approximately ballistic trajectories on the length scale of the simulation domain and have no trouble penetrating into the passive space, which itself is slowly dispersed as the passive objects diffuse with $\Davp$, induced by the hydrodynamic interactions. At higher volume fractions, the phase-separation is destroyed very rapidly, on the order of a few time units, as near-field hydrodynamic and steric interactions displace passive particles.
	
	The exception to this behaviour are neutral and puller suspensions ($\beta\geq0$) at very high packing density $\phi=0.5$, and low-to-moderate active fractions $\alpha\leq 0.5$ (Figure~\ref{fig4}a-c and supplementary movies 4-6). Here, we observe the formation of a coherent state in the active layer, with a direction of motion parallel to the active-passive interface. In the case of neutral squirmers ($\beta=0$) there is no attraction at leading order between the two types of objects, so the interface remains almost entirely undisturbed once the coherent motion has emerged. For pullers ($\beta=1$), there is lateral repulsion present (Figure~\ref{fig:illust}), and comparing the two cases in motion, it is apparent that the interface is perturbed but held together by steric interactions between the passive particles. When the active fraction dominates, $\alpha= 5/6$, the passive layer becomes too thin for the internal steric interactions to hold it together, and we recover behaviour of the initially mixed suspension. Likewise, just a slight increase of the average inter-particle distance by $7.7\%$ (equivalent to reducing the volume fraction $\phi$ from $0.5$ to $0.4$) makes the effect disappear.
	
	This effect can be quantified using the spatial cross-correlation $S$ that we introduced in section \S\ref{sec:metrics}. In Figure~\ref{fig4}d-f, we illustrate the particle densities
	defined by projecting in a plane orthogonal to the plane of initial phase-separation, and the associated correlation $S$. As expected, $S$ is more strongly negative the more well-defined the phase-separation between swimmers and passive spheres is. In Figure~\ref{fig4}g we compare the value of $S$ at $t=180$ for the two different initial conditions at high densities ($\phi= 0.5$). For
	pushers, we find that $S$ is close to $0$ regardless of the condition, indicating a homogeneous mix of the two particles. For neutral swimmers, we detect a strong difference as described above. For pullers, the difference is again small, although $S$ is more negative than for
	pushers, indicating a tendency of pullers to create weakly defined clusters of passive particles. 
	
	Taken together with the findings for an initially mixed suspension, it may therefore be said that the formation of a coherent structure is a necessary but not a sufficient condition to preserve phase-separation. Additionally, the suspension needs to be very dense, and the proportion of active swimmers must not be higher than $0.5$.
	Furthermore, since there is no indication that the phase-separation emerges dynamically from an initially mixed suspension, it should be interpreted as a metastable state that depends on the initial configuration.
	We note that \cite{zhou2025} investigated the behaviour of squirmers within a film and demonstrated that hydrodynamic interactions disrupt phase separation. Whilst our study discusses the separation of active and passive particles, focusing solely on active particles reveals no phase separation. This result does not contradict the assertions of \cite{zhou2025}.

	\subsection{Weak bottom-heaviness ($\gbh=10$): phase-separation and particle transport}
	
	\begin{figure}[p!]
		\centering
		\centering
		\includegraphics[width=\textwidth]{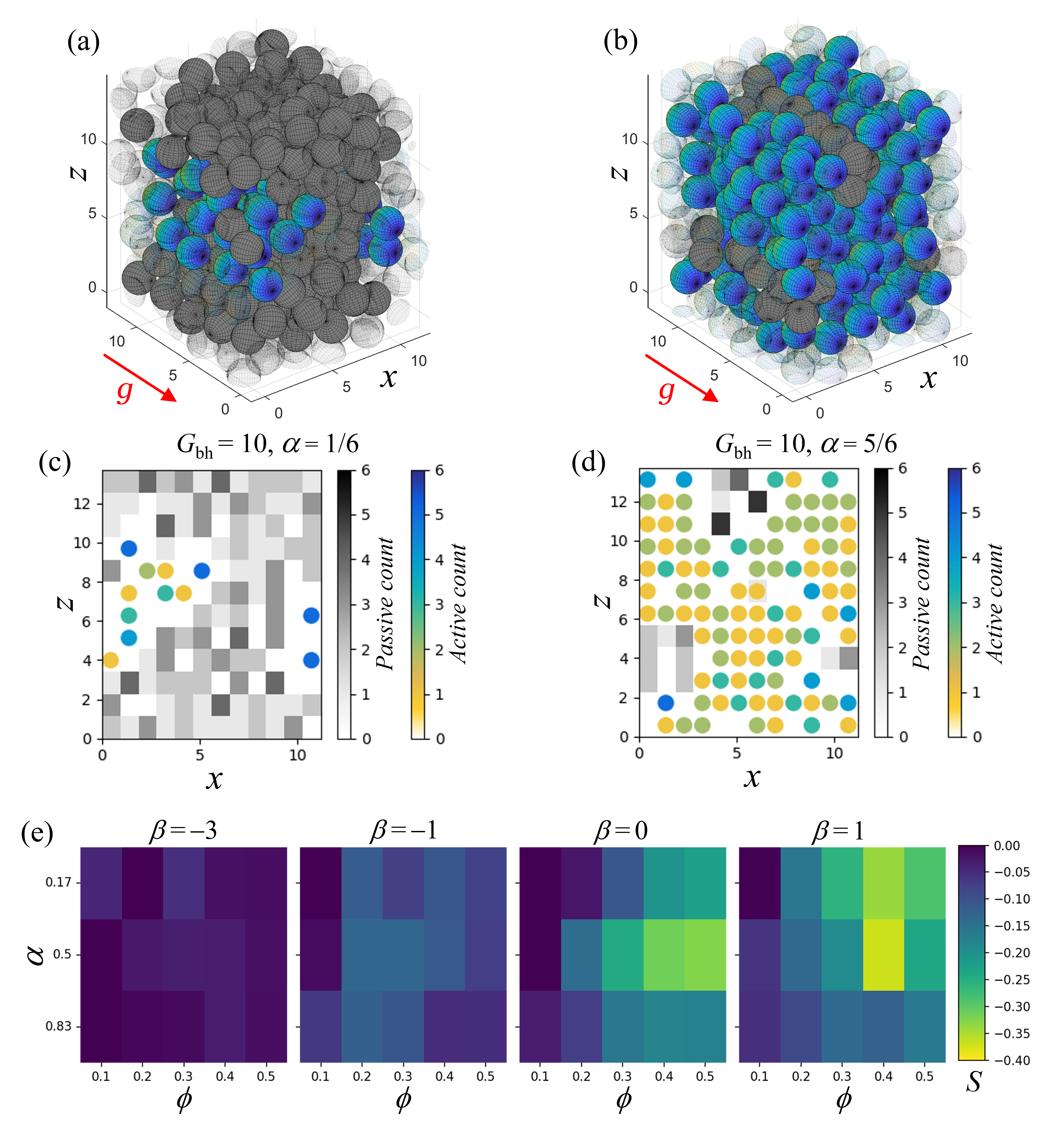}
		\caption{Phase-separation of initially mixed suspensions in the case of
			weak bottom-heaviness ($\gbh=10$).
			(a,b) Illustration of the steady state $(t\approx 180)$ with (a) $\alpha=1/6$ and (b) $\alpha=5/6$ ($\phi=0.5$, and $\beta=0$). See also supplementary movies 7-8.  Irrespective of the ratio of particle types, lanes of swimmers form that separate from passive particles in the plane orthogonal to the direction of alignment ($-\bm{g}$, in red). 
			(c,d) Density projection into a plane orthogonal to the direction of gravity, and sphere counts in a 2D grid of bins with (c) $\alpha = 1/6$ and (d) $\alpha = 5/6$ ($\phi=0.5$, and $\beta=0$). The spatial cross-correlations are $S = -0.23$ for $\alpha = 1/6$ and $S = -0.17$ for $\alpha = 5/6$.
			(e) Heatmap of the steady-state cross-correlation $S$ for weakly bottom-heavy suspensions. Larger negative values indicate more pronounced phase-separation.}
		\label{fig5}
	\end{figure}

	We now consider the case in which the active microswimmers are weakly bottom-heavy, as would be the case for the green alga \emph{Volvox} \citep{Drescher2009,Ishikawa_Pedley_2020}. This breaks the symmetry of the system by introducing a preferred orientation in the direction opposed to gravity, $-\bm{g}$. In pure squirmer suspensions, this is known to lead to a number of interesting effects such as gravitaxis, plume formation, and the emergence of orientational order in pusher suspensions that would otherwise not be stable \citep{ishikawa2008development,ishikawa2022instability,Ishikawa_Brumley_Pedley_2025}.
	
	In our simulations we find that this generally holds true also in the presence of passive particles. We note that only the initially mixed suspensions are considered in this section.
	When only a few passive objects are present ($\alpha=5/6$), the picture is very similar to a pure squirmer suspension ($\alpha=1$). Here, bottom-heaviness is generally sufficient to enforce almost complete alignment except in the case for very strong pushers ($\beta=-3$), where it is reduced by up to $||\langle \bm{e} \rangle_a||\approx 0.5$ for very dense suspensions.  When more obstacles are introduced, ($\alpha\leq 0.5$), the presence of the passive particles has a visible effect at high packing densities ($\phi\geq 0.4$). In the case of the more unstable pushers, the emergence of orientational order is strongly suppressed compared to the case without obstacles. In the case of neutral squirmers, the emergence is merely delayed. In the case of pullers, it is delayed and partially suppressed.
	
	This difference in behaviour may be understood by considering the different types of effect the active swimmers have on neighbouring obstacles (Figure~\ref{fig:illust}). 
	Previous studies have shown that when strong pushers approach obstacle particles, they may undergo significant directional changes due to lubrication forces and become trapped \citep{chamolly2017active,Ishikawa-JAP2019}. Such large directional changes make the ordered structure more susceptible to collapse.
	On the other hand, these previous studies have reported that neutral swimmers and pullers are not captured by the obstacle particles and swim relatively straight through the lattice arrangement of obstacle particles. This straight-line swimming behaviour contributes to maintaining the orientational order.
	This leads to a sorting and phase-separation of the suspension into ``lanes'' of aligned swimmers and a nearly stationary passive phase between them (Figure~\ref{fig5}a-d and supplementary movies 7-8). This sorting already occurs at lower passive fractions, but the time it takes for the separation to occur increases with the proportion of passive obstacles.
	However, similarly to the phase-separated case with $\gbh=0$, the stronger hydrodynamic interactions between the phases disturb this process at very high densities. As a result, the pullers fail to organise into a state with well-defined translational symmetry along the axis of gravity. This explains the joint delay and reduction of orientational order compared to the case with no obstacles for pullers.
	
	In Figure~\ref{fig5}e we illustrate this effect quantitatively with the cross-correlation $S$, obtained by projecting in the direction of gravity. For pushers ($\beta<0$), or a small fraction of passive spheres ($\alpha=5/6$) there is no phase-separation. For neutral squirmers ($\beta=0$), the effect increases with packing density and is most pronounced for an equal mix of particle types. For pullers ($\beta=1$), phase-separation is maximised at a packing density of $\phi= 0.4$, but increased interactions at even higher densities lead again to more mixing.
	
	\begin{figure}[t!]
		\centering
		\includegraphics[width=0.8\columnwidth]{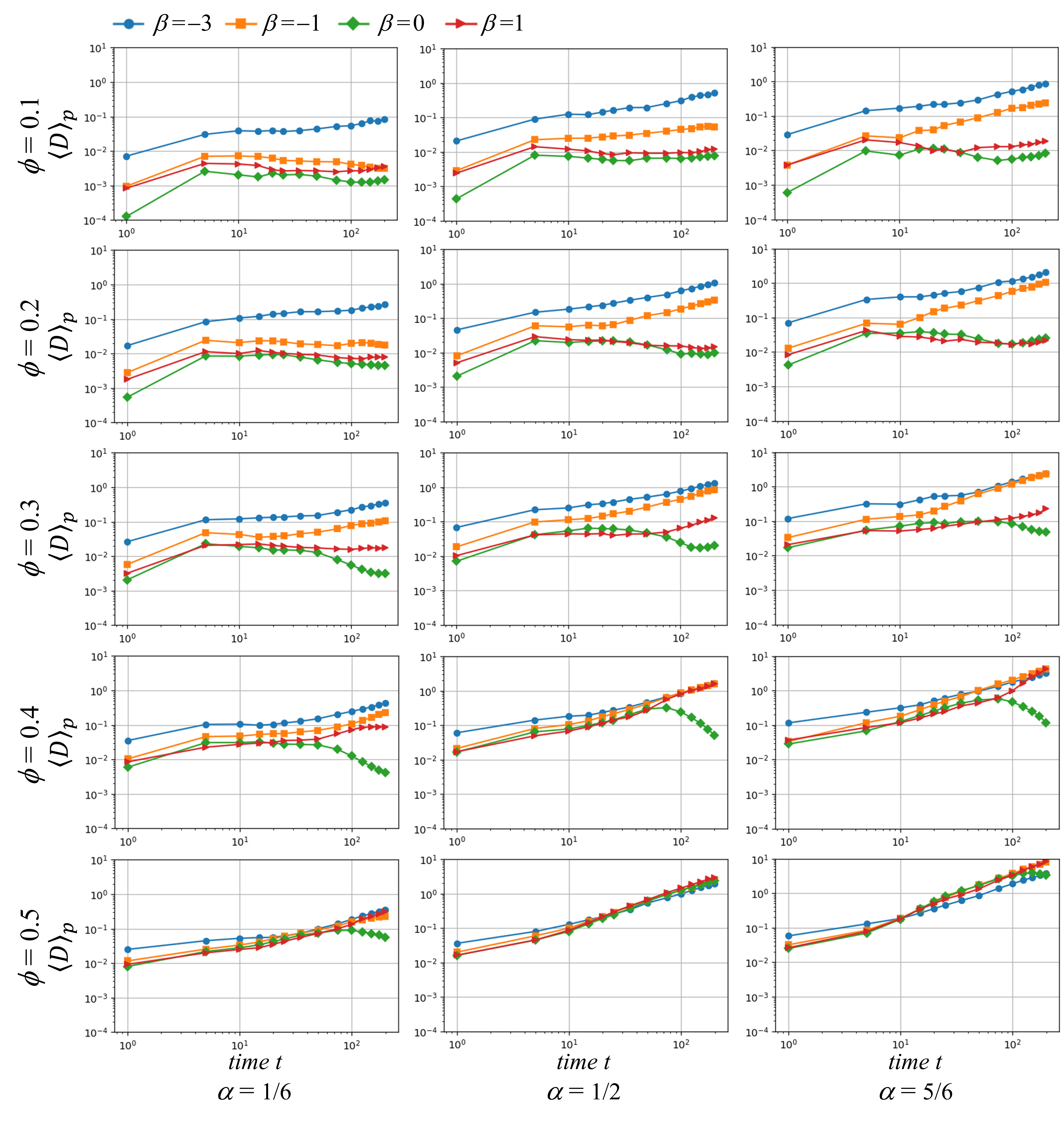}
		\caption{Temporal evolution of the dispersion coefficient of passive particles $\Davp$ for suspensions of weakly bottom-heavy squirmers ($\gbh=10$) in an initially mixed state with random orientations.}
		\label{fig:bh10diff}
	\end{figure}
	
	We previously established that in the absence of bottom-heaviness, the passive particles are poorly mixed, but experience some ballistic transport when an ordered phase exists. With bottom-heaviness, this effect is generalised due to the more ubiquitous and stable nature of this ordered state, but it is not congruent in parameter space with the phase-separation. At low active fraction ($\alpha= 1/6$), $\Davp$ continues to be extremely small (Figure~\ref{fig:bh10diff}). However, as $\alpha$ is increased, we again observe that the passive particles are transported ballistically by the active ones. Interestingly, we observe that at lower packing densities ($\phi\leq 0.3$), pushers are the most efficient at transporting
	despite their tendency to destabilise the ordered state through their hydrodynamic interactions.
	This is because these interactions also lead to a lateral attraction between pushers and their surroundings, which is beneficial to transport when order is enforced externally through bottom-heaviness. At high packing densities ($\phi\geq 0.4$), pullers are as efficient as weak pushers for transport, while strong pushers ($\beta=-3$) lose their advantage due to their reduced orientational order. Neutral swimmers are only transiently efficient at transporting, and cease to affect the passive particles once phase-separation has manifested.
	
	\subsection{Strong bottom-heaviness ($\gbh=100$): coherent and sandwich-like structures}
	\begin{figure}[p!]
		\centering
		\includegraphics[width=\textwidth]{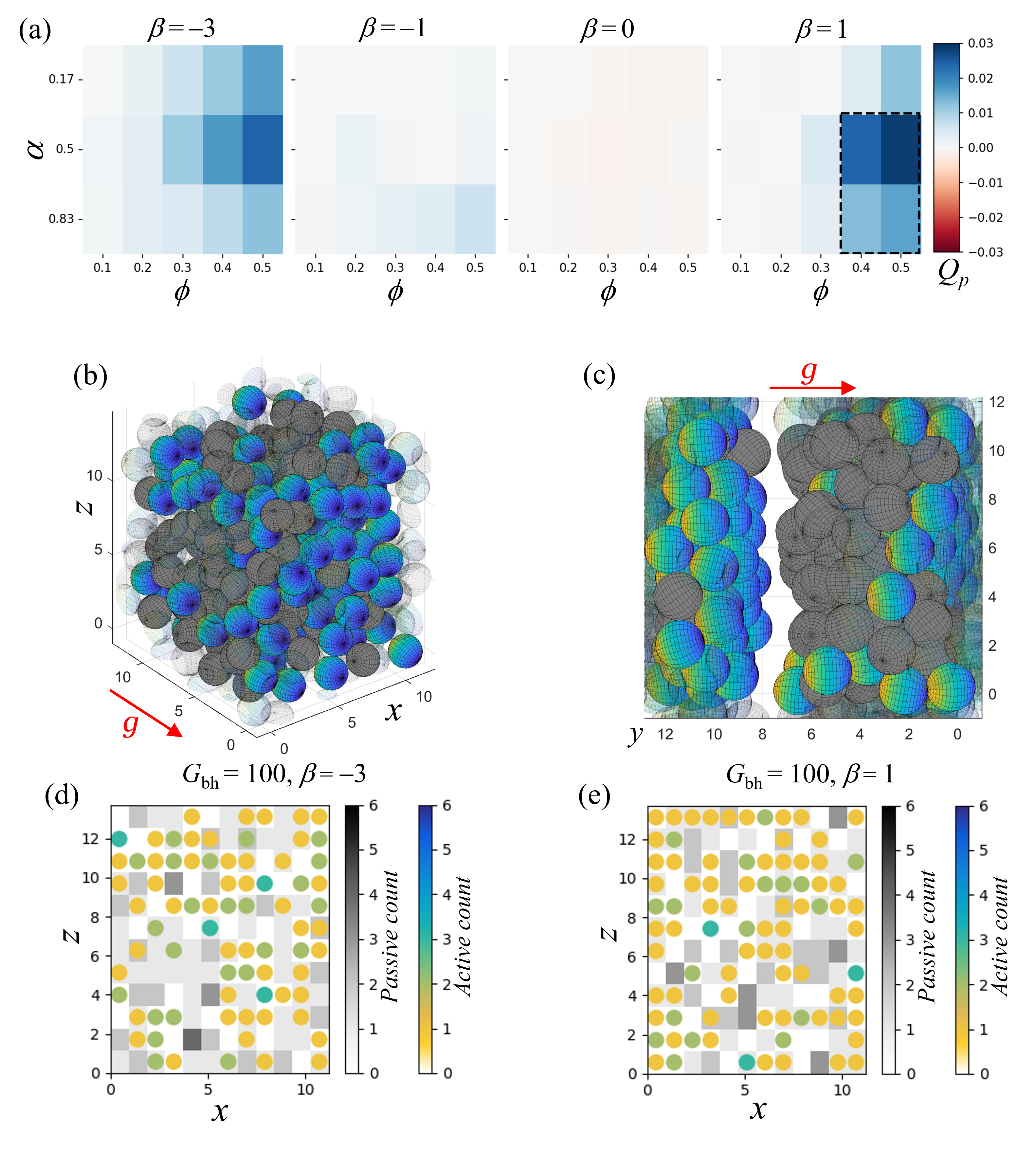}
		\caption{Examples of the transport of passive particles in the presence of strong bottom-heaviness with $\gbh=100$.
			(a) Flux of passive particles in the direction of alignment, $Q_p$, in a mixed suspension with strongly bottom-heavy squirmers. The dashed outline indicates the region in parameter space in which lamellar phase-separation occurs.
			(b,c) Illustration of steady state with (b) $\beta = -3$ and (c) $\beta = 1$ ($\phi=0.5$, $\alpha = 1/2$). See also supplementary movies 9-10. The direction of alignment is $-\bm{g}$, shown in red.
			Strongly directed pushers are able to maintain a coherent formation and transport passive objects through lateral hydrodynamic attraction (c.f.\ Fig.~\ref{fig:illust}). On the other hand, strongly directed pullers form a sandwich-like structure at high densities, with a layer of passive particles pushed by a layer of swimmers, followed by a gap.
			(d,e) Density projection into a plane orthogonal to the direction of gravity, and sphere counts in a 2D grid of bins with (d) $\beta = -3$ and (e) $\beta = 1$ ($\phi=0.5$, $\alpha = 1/2$).}
		\label{fig7}
	\end{figure}
	
	Finally, we consider the case of very strong bottom-heaviness. In this case, the active swimmers are essentially locked in their orientation along the negative axis of gravity. As such, a nearly perfectly ordered state is enforced for all squirmer types, active fractions and packing densities. We note again that, in this section, only the initially mixed suspensions are considered.
	
	Across the range we note a strong, approximately proportional, increase in $\Uavp$ and $\Davp$ with active fraction $\alpha$. In other words, when fewer passive particles are present, these are transported more efficiently. This means that the flux of passive particles $Q_p= N_p \langle - \hat{\bm{g}} \cdot \bm{U} \rangle_p \times\phi/ \tfrac{4N}{3}\pi a^3$ is optimal for an equal mix of particle types (Figure~\ref{fig7}a). Similarly to the case of weak bottom-heaviness, we find that strong pushers ($\beta=-3$) are efficient at transporting passive particles due to the strong lateral attraction (Figure~\ref{fig:illust}). Strongly directed pushers are able to maintain a coherent formation, and the two phases are moving while mixing (Figure~\ref{fig7}b,d and supplementary movie 9).
	For neutral squirmers ($\beta=0$) we again observe the fibrillar (to the gravitational axis) phase-separating effect and correspondingly transient ability to transport. Additionally, we observe this effect now with a slower onset for weak pushers ($\beta=-1$).
	
	For pullers ($\beta=1$), we find a new behaviour that is unique to this strongly aligned type of suspension. At high packing densities $\phi\geq 0.4$ and active fractions $\alpha\geq 0.5$ we observe a different kind of phase-separation, with a 3-layered structure of passive obstacles pushed by active swimmers, followed by an empty fluid region (Figure~\ref{fig7}c,e and supplementary movie 10). This new type of lamellar phase-separation arises from three different effects. The dipolar flow of the pullers attracts other pullers from the front and back, resulting in a horizontal layer of pullers.
	Leftward and rightward displacements are stabilised by repulsive forces from left and right neighbours, whereas upward and downward perturbations are amplified by attractive forces from the top and bottom neighbours \citep{Lauga_2021}.
	The dipolar flow also directs passive particles to position themselves either in front of or behind a squirmer (Figure~\ref{fig:illust}). The strong bottom-heaviness ensures that this process always occurs in the same direction and without any reorientation of squirmers, which explains the longitudinal nature of the effect. Passive spheres in the wake will remain approximately stationary until they are reached by the next wave of periodically arriving particles, leading to an accumulation of passive objects in front of the squirmers. This layer is then stabilised by steric interactions between passive particles that prevent all but the occasional escape of squirmers through the passive layer, which explains why this effect again occurs only at very high packing densities $\phi$ of the suspension.

	\section{Discussion}
	\label{discuss}
	
	In this paper we analysed the effect of neutrally buoyant passive obstacles on a 3D suspension of squirmers. We examined a wide parameter space of different swimmer types, packing densities, proportion of active and passive particles, bottom-heaviness, and initial conditions. We found that in the absence of an external director field, such as gravity, known results about the collective behaviour of squirmer suspensions are reproduced qualitatively and quantitatively when the proportion of passive spheres is small. Generally, the more passive spheres are present, the more the formation of orientational order is disturbed. As in pure squirmer suspensions, alignment correlation is always a global effect, with the average alignment correlation depending only very weakly on interparticle separation. At very high packing densities, some behaviour special to mixed suspension emerges, with an initially phase-separated suspension of neutral squirmers or pullers remaining phase-separated with a collective state emerging within the quasi-2D confinement due to the passive obstacles. Once bottom-heaviness is introduced, phase-separation occurs dynamically in some cases, notably into a fibrillar structure for neutral squirmers and pullers at medium densities, and a lamellar structure for pullers at high densities with very strong external alignment. 
	
	Our results can be understood well in the context of the microstructures observed in previous studies.
	The orientational ordering of squirmers in pure 3D suspensions, first reported by \cite{ishikawa2008development}, was also observed presently for neutral and puller ($\beta = 1$) squirmers in pure active suspensions (see Figure \ref{fig3}). However, when passive spheres are introduced, the orientational order is weakened. This can be explained by near-field hydrodynamic interactions between particles. In Stokes flow, solutions can be superimposed, so the interaction between squirmer A and squirmer B can be expressed as the superposition of the interaction between squirmer A and a passive sphere and the interaction between squirmer B and a passive sphere \citep{Ishikawa2024ARFM}. Therefore, when one particle becomes a passive sphere, the effect of the squirming velocity is halved, and the effect of aligning the direction is also halved. This reduced alignment leads to weakened orientational order.
	
	Introducing bottom-heaviness caused phase-separation into a fibrillar structure (see Figures \ref{fig5} and \ref{fig7}), which can be considered like a jet of a squirmer suspension.
	\cite{ishikawa2022instability} investigated the instabilities of a jet of a suspension of bottom-heavy squirmers. They found
	that the jets are unstable, with different unstable modes for pullers and pushers. The jet of pullers breaks up into droplets, while the jet of pushers buckles and undergoes a waving instability. In the present study, however, such instabilities in the jet were not clearly observed. For pushers, HIs with passive particles are universally too strong for a jet to develop, much less be stable. The sandwich-like phase separation for very bottom-heavy pullers 
	is reminiscent of the droplet breakup, however a very strong external orienting torque
	suppresses the formation of spherical drops, stabilising them in a layered state. Otherwise, the passive spheres surrounding the active jets, or lanes, form a wall-like structure that suppresses the instability. Therefore, the jet structure is more stable in mixed than in pure puller suspensions.
	
	Using the same parameters ($\beta = 1, \gbh = 100$), \cite{ishikawa2008coherent} also observed a horizontal band of puller squirmers in a quasi-2D pure suspension. The band formation was also observed by \cite{Spormann1987} in experiments using magnetotactic bacteria.
	It is intriguing that similar horizontal structures appeared in both 2D and 3D analyses and experiments.
	This microstructure appears because puller squirmers attract other squirmers in front of and behind them, creating a flow that pushes them sideways.
	However, in pure 3D suspensions of puller squirmers, the horizontal band has not been observed \citep{ishikawa2008development}.
	The lamellar kind of phase-separation only appeared when passive particles were mixed in ($\alpha = 1/2$), indicating that the microstructure of a mixed suspension can differ qualitatively from that of a pure suspension.
	
	The instability of horizontal layers presents a highly intriguing topic. \cite{Eisenmann_2025} experimentally measured the growth rate of the puller horizontal band instability, demonstrating its development within approximately 30 seconds. A similar instability was analysed in the vertical plume of the pusher squirmer \citep{ishikawa2022instability}, because the stability of a horizontally arranged pullers is analogous to that of a vertically arranged pushers.
	In this setting, the instability developed over approximately 20 non-dimensional time units for squirmers with $\beta = 1$. In contrast to these timescales, the current study performs computations at a non-dimensional time of 200, which is ten times longer. This timescale is considered sufficient for discussing the instability. However, the wavelength of the band instability is approximately four times the band width \citep{Marco_2025}. The domain size used in the current study is not sufficiently long relative to this, necessitating discussion of the instability within a larger computational domain. This remains a future challenge. Meanwhile, a study discussing instability in a single line of pusher squirmers identified swimmer-level instability where swimmers shift alternately, i.e. zigzag instability \citep{Lauga_2021}. In the sandwich-like phase separation of the puller shown in Fig. \ref{fig7}c, numerous locations exhibit adjacent swimmers that are shifted fore and aft. However, these shifts do not grow due to interference from passive particles. It is considered that passive particles inhibit the growth of instability.
	
	In general, the transport of passive particles is very weak when there is no external director field, and only becomes important in a region of parameter space where there is no orientational order. When an external director is introduced, pushers are somewhat effective at transporting passive particles because the external field introduces order in a regime with strong HIs, and this effect becomes more pronounced when the alignment is strong. Neutral squirmers are poor in this respect, since their weak interactions favour phase-separation into a free-flowing active and near-stationary passive phase. Pullers have the most complex characterisation, with stable alignment 
	whilst exhibiting hydrodynamic interactions that displace passive particles laterally and do not promote passive transport
	except in the extreme case of lamellar phase-separation for the most dense and strongly aligned suspensions.
	
	This paper analysed model swimmers and particle suspensions, but similar situations could realistically occur. Many microalgae exhibit gravitaxis, responding to gravity, and various swimming modes exist, such as pull-type (ex. \textit{Chlamydomonas} \citep{Goldstein2015}), neutral-type (ex. \textit{Volvox} \citep{Goldstein2015}), and push-type (ex. \textit{Prymnesium} \citep{Dolger2017}). Situations where such microalgae coexist with non-motile microorganisms or with solid suspended particles are considered to be close to the conditions envisaged in this study. Whilst gravitational torque was considered here, the form of the fundamental equations remains similar for magnetic torque. Therefore, scenarios where magnetic microswimmers transport other particles or operate in complex environments are also considered close to the conditions envisaged in this study. Furthermore, even in cases of biological taxis, where the microorganism itself initiates directional change rather than responding to an external torque, the resulting motion patterns are anticipated to be somewhat similar. This scenario encompasses various situations, such as bacteria exhibiting chemotaxis in granular media or microalgae exhibiting phototaxis in particle suspension. The results of this study should also aid in understanding the outcomes of such experiments.

	The results obtained in this study indicate that microstructure and particle transport undergo significant changes depending on the type of swimmer, suggesting the importance of hydrodynamic interactions.
	The obtained knowledge is important for understanding the behaviour of active particles in complex fluids and for controlling them using external torques.

	\appendix
	\section{Numerical method}\label{appendix:numericalmethod}
	In this paper we employ numerical code that was originally developed to model the dynamics of a suspension of squirmer microswimmers. Here we give a brief overview of the procedure and how we adapted it. The full mathematical details may be found in \cite{ishikawa2008development} and \cite{ishikawa2012vertical}.
	
	In essence, the default mode of calculating the interaction between the
	squirmers and the obstacle particles is by means of a far-field expansion, which is computed using Ewald summation. In the event that the distance between the 
	particles falls below a predetermined threshold, the pairwise interaction is analysed by lubrication theory and computed using a precompiled database of lubrication interactions.
	
	The fluid velocity $\bm{u}(\bm{x})$ at any point external to the swimmer or the lattice may be written in terms of a boundary integral as \citep{pozrikidis1992boundary}
	\begin{align}
		\bm{u}(\bm{x})-\langle\bm{u}\rangle =-\frac{1}{8\pi\mu}\sum_{\text{spheres}}\int_S \bm{J}(\bm{x}-\bm{y})\cdot\bm{q}(\bm{y})\,\text{d}S,
	\end{align}
	where $\langle\bm{u}\rangle$ is the average flow velocity in the computational domain, the sum is performed over all 216 spheres within and the integral is calculated over the surface of each sphere. The integrand consists of $\bm{J}$, which is the Green's function for a triply periodic lattice of spheres, and $\bm{q}$, which is the single-layer potential. At each time step $\bm{J}$ is calculated by Ewald summation as detailed in \cite{beenakker1986ewald}. The right-hand side may be expanded in moments about the centre of each sphere as
	\begin{align}\label{act:eq:expansion}
		\bm{u}(\bm{x})-\langle\bm{u}\rangle=\frac{1}{8\pi\mu}\sum_{\text{sphere centres}}\left[\text{ monopole of }\bm{q}+\text{ dipole }+\text{ quadrupole }+\dots\right],
	\end{align}
	where each term is contracted with a propagator that is calculated from $\bm{J}$ and its gradients. The monopole and antisymmetric dipole correspond to the force and torque on each sphere respectively.
	
	In order to determine the swimmer kinematics, Fax\'en's laws are applied for each sphere $i$ located at $\bm{x}_i$ in the domain as
	\begin{align}\label{act:eq:faxen1}
		\bm{U}_i-\langle\bm{u}\rangle &=\frac{\bm{F}_i}{6\pi\mu a} + U_0 \bm{e} +\left(1+\frac{a^2}{6}\nabla^2\right)\bm{u}'(\bm{x}_i),
		\\
		\bm{\Omega}_i-\langle\bm{\omega}\rangle &=\frac{\bm{G}_i}{8\pi\mu a^3}+\frac{1}{2}\bm{\nabla}\times\bm{u}'(\bm{x}_i),\label{act:eq:faxen2}\\
		-\langle\bm{E}\rangle &=\frac{\bm{S}_i}{\tfrac{20}{3}\pi\mu a^3} + \frac{3}{10} \mu a^2 U_0 \beta\left(3\bm{e}\bm{e}-\bm{I}\right) + \frac{1}{2\mu}\left(1+\frac{a^2}{10}\nabla^2\right)\bm{\sigma}^{\prime}_{\text{dev}}(\bm{x}_i),\label{act:eq:faxen3}
	\end{align}
	where $\bm{U}_i$, $\bm{\Omega}_i$ are the translational and rotational velocity of the $i$th sphere, $\bm{F}_i$, $\bm{G}_i$ and $\bm{S}_i$ are the force, torque and stresslet that the $i$th sphere exert on the fluid, $\bm{u}'$ and $\bm{\sigma}^{\prime}_{\text{dev}}$ are the velocity and deviatoric stress due to the other spheres. $\langle\bm{u}\rangle$, $\langle\bm{\omega}\rangle$ and $\langle\bm{E}\rangle$ are the domain-averaged velocity, vorticity and rate of strain, respectively. The second terms in the right-hand side of Eqs. \eqref{act:eq:faxen1} and \eqref{act:eq:faxen3} are present only for the squirmer, which has the orientation vector $\bm{e}$.
	
	The method then proceeds by substituting Eq.~\eqref{act:eq:expansion} into the right-hand side of Eqs.~\eqref{act:eq:faxen1}-\eqref{act:eq:faxen3} and truncating the expansion after the dipole.
	Furthermore, the equations are nondimensionalised by using the particle radius $a$, the swimming speed of a solitary squirmer $U_0$, and the viscosity $\mu$.
	This leads to a linear system of the form
	\begin{align}\label{act:eq:farfield}
		\begin{pmatrix}
			\bm{U}-\langle\bm{u}\rangle -\bm{e}\\
			\bm{\Omega}-\langle\bm{\omega}\rangle\\
			-\tfrac{3}{10}\beta(3\bm{e}\bm{e}-\bm{I})
		\end{pmatrix}
		=
		\bm{M}^{\text{far}}\cdot
		\begin{pmatrix}
			\bm{F}\\
			\bm{G}\\
			\bm{S}
		\end{pmatrix},
	\end{align}
	where $\bm{M}^{\text{far}}$ is the far-field grand mobility matrix, which is the same as the one for inert spheres \citep{brady1988stokesian}.  The infinite extent of the suspension is taken into account using Ewald summation \citep{beenakker1986ewald}.
	
	As in \cite{ishikawa2012vertical}, the lubrication force between
	squirmers in close proximity is introduced in a pairwise additive fashion by the following equation:
	\begin{eqnarray}
		\label{mat.5}
		\left[ \bm{I} + \bm{M}_{FU}^{far} : \bm{K}_{FU}^{2b} \right] .
		\left( \begin{array}{c}
			\bm{U} - \langle \bm{u} \rangle -\bm{e} \\
			\bm{\Omega} - \langle \bm{\omega} \rangle
		\end{array} \right)
		= ~~~~~~~~~~~~~~~~~~~~~~~~~~~~~~~~~~~~~~\\
		\nonumber
		\bm{M}_{FU}^{far} .
		\left\{
		\left( \begin{array}{c}
			\bm{F} + \bm{F}^\alpha \\
			\bm{G} + \bm{G}^\alpha
		\end{array} \right)
		- \bm{K}_{SU}^{2b} :
		\left[ - \langle \bm{E} \rangle
		- \frac{3}{10} \beta \left( 3 \bm{e} \bm{e} - \bm{I} \right)
		\right]
		\right\}
		+ \bm{M}_{SU}^{far} : \bm{S}^{far}~,
	\end{eqnarray}
	with
	\begin{equation}
		\left( \begin{array}{c}
			\bm{F}^\alpha \\
			\bm{G}^\alpha
		\end{array} \right)
		=
		\bm{K}_{2B}^{near} :
		\left( \begin{array}{c}
			-\bm{e} \\
			0 \\
			- \frac{3}{10} \beta \left( 3 \bm{e} \bm{e} - \bm{I} \right)
		\end{array} \right)
		-
		\left( \begin{array}{c}
			\bm{F}_{sq}^{near} \\
			\bm{G}_{sq}^{near}
		\end{array} \right) ,
	\end{equation}
	and
	\begin{equation}
		\bm{M}^{far} = 
		\left[ \begin{array}{cc}
			\bm{M}_{FU}^{far} & \bm{M}_{FE}^{far} \\
			\bm{M}_{SU}^{far} & \bm{M}_{SE}^{far}
		\end{array} \right]
		~,~~
		\bm{K}^{2b} = 
		- \bm{K}_{2B}^{far} + \bm{K}_{2B}^{near}
		=
		\left[ \begin{array}{cc}
			\bm{K}_{FU}^{2b} & \bm{K}_{FE}^{2b} \\
			\bm{K}_{SU}^{2b} & \bm{K}_{SE}^{2b}
		\end{array} \right]~,
		\label{split}
	\end{equation}
	where $\bm{S}^{far}$ is the far-field contribution to the stresslet, which is approximated by neglecting the additional contribution of cell-cell interactions. $\bm{K}_{2B}^{far}$ is the far-field two-body resistance matrix, and $\bm{K}_{2B}^{near}$ is the near-field two-body resistance matrix. The mobility and resistance matrices are split into four components, as in Eq.(\ref{split}): subscripts $FU, FE, SU$, and $SE$ indicate coupling between force and velocity, force and rate of strain, stresslet and velocity, and stresslet and rate of strain, respectively. $\bm{F}_{sq}^{near}, \bm{G}_{sq}^{near}$ and $\bm{S}_{sq}^{near}$ are respectively the force, torque and stresslet generated by the two-squirmer interaction in the near-field, calculated numerically using the boundary element method \citep{ishikawa2006hydrodynamic}. 	
	Finally, a very short range repulsive force is additionally included to model steric interactions and regularise the computational time step \citep{ISHIKAWA_PEDLEY_2007}.
	
	\section{Properties of the $S$-operator}\label{appendix:sproperties}
	
	Here we show that $S^2=1$ as claimed. Roughly speaking, this is true because the $\bm{z}$ are normalised, $W$ is an averaging operator, and the Cauchy-Schwarz inequality applies to the inner product that defines $S$. We define $M=m_im_j$ as the number of grid points on which the matrices $\bm{z}_{a,p}$ are computed, and flatten them into vectors of length $M$. Since they are normalised by the mean and standard deviation over the grid, we have
	\begin{align}
		||\bm{z}_{a.p}||^2 = M.
	\end{align}
	Similarly, the tensor $W$ can be flattened into a matrix operating on these flattened vectors. By construction, this matrix is symmetric and each row sums to 1, it is therefore a stochastic matrix. From this it follows that its eigenvalues are real and satisfy $|\lambda| \leq 1$. This in turn implies that
	\begin{align}
		||W\bm{v}||^2 \leq ||\bm{v}||^2,
	\end{align}
	for any vector $\bm{v}$.
	
	The definition of $S$ can hence be written as
	\begin{align}
		S = \frac{1}{M}\bm{z}_a^TW\bm{z}_p = \frac{1}{M} \langle \bm{z}_a, W\bm{z}_p \rangle,
	\end{align}
	where double square brackets indicate the inner product. Applying the Cauchy-Schwarz inequality to this inner product yields the desired result:
	\begin{align}
		S^2 = \frac{1}{M^2} | \langle \bm{z}_a, W\bm{z}_p \rangle |^2 \leq \frac{1}{M^2} ||\bm{z}_a||^2  ||W\bm{z}_p||^2\leq \frac{1}{M^2} ||\bm{z}_a||^2  ||\bm{z}_p||^2 = 1.
	\end{align}

	\section{Supplementary movie captions}\label{appendix:moviecaptions}
	
	\textbf{Note: the videos are available in the officially published open access version of this article \cite{chamolly2026self}. Please note that due to an error in the editorial process, the links to the videos are out of order (1, 10, 2-9).}
	
	All supplementary movies have the same format. On the left side, a movie of the simulation is played.  Active spheres are in colour, and passive spheres in grey. Periodic copies of spheres are transparent. On the top right, the mean velocity of active and passive spheres, $U_a$ and $U_p$, as well as the scalar magnitude of the mean orientation vector of the active spheres $\langle \bm{e} \rangle_a$ are plotted as a function of simulation time. On the bottom right, the scalar dispersions $\langle D \rangle_a$, $\langle D \rangle_p$ are plotted as a function of time.\\
	
	\textbf{Supplementary movie 1:} (cf.\ Fig.~\ref{fig3}a) Dynamics of an initially mixed suspension with active proportion $\alpha=1/6$ of pullers ($\beta=1$), at packing density $\phi=40\%$ and no bottom-heaviness ($G_\text{bh}=0$). No orientational order develops. \\
	
	\textbf{Supplementary movie 2:} (cf.\ Fig.~\ref{fig3}b) Dynamics of an initially mixed suspension with active proportion $\alpha=1/2$ of pullers ($\beta=1$), at packing density $\phi=40\%$ and no bottom-heaviness ($G_\text{bh}=0$). Almost no orientational order develops. \\
	
	\textbf{Supplementary movie 3:} (cf.\ Fig.~\ref{fig3}c) Dynamics of an initially mixed suspension with active proportion $\alpha=5/6$ of pullers ($\beta=1$), at packing density $\phi=40\%$ and no bottom-heaviness ($G_\text{bh}=0$). A state of orientational order develops, but the direction of order is not stable. \\
	
	\textbf{Supplementary movie 4:} (cf.\ Fig.~\ref{fig4}a) Dynamics of an initially phase-separated suspension with active proportion $\alpha=1/6$ of neutral squirmers ($\beta=0$), at packing density $\phi=50\%$ and no bottom-heaviness ($G_\text{bh}=0$). Phase-separation is maintained and a coherent state of motion develops in the active phase. \\
	
	\textbf{Supplementary movie 5:} (cf.\ Fig.~\ref{fig4}b) Dynamics of an initially phase-separated suspension with active proportion $\alpha=1/2$ of neutral squirmers ($\beta=0$), at packing density $\phi=50\%$ and no bottom-heaviness ($G_\text{bh}=0$). Phase-separation is maintained and a coherent state of motion develops in the active phase. \\
	
	\textbf{Supplementary movie 6:} (cf.\ Fig.~\ref{fig4}c) Dynamics of an initially phase-separated suspension with active proportion $\alpha=5/6$ of neutral squirmers ($\beta=0$), at packing density $\phi=50\%$ and no bottom-heaviness ($G_\text{bh}=0$). The phase-separation is not stable, but a coherent state develops with a stable orientation, in which the passive particles are transported. \\
	
	\textbf{Supplementary movie 7:} (cf.\ Fig.~\ref{fig5}a) Dynamics of an initially mixed suspension with active proportion $\alpha=1/6$ of neutral squirmers ($\beta=0$), at packing density $\phi=50\%$ and moderate bottom-heaviness ($G_\text{bh}=10$). A fibrillar state of phase separation develops with linear jets of squirmers piercing through a nearly stationary passive environment.\\
	
	\textbf{Supplementary movie 8:} (cf.\ Fig.~\ref{fig5}b) Dynamics of an initially mixed suspension with active proportion $\alpha=5/6$ of neutral squirmers ($\beta=0$), at packing density $\phi=50\%$ and moderate bottom-heaviness ($G_\text{bh}=10$). A fibrillar state of phase separation develops with isolated threads of passive particles in an otherwise unperturbed coherently moving active phase.\\
	
	\textbf{Supplementary movie 9:} (cf.\ Fig.~\ref{fig7}b) Dynamics of an initially mixed suspension with active proportion $\alpha=1/2$ of strong pushers ($\beta=-3$), at packing density $\phi=50\%$ and strong bottom-heaviness ($G_\text{bh}=100$). Passive particles are transported in the direction of the director field through strong hydrodynamic interactions with the pullers.\\
	
	\textbf{Supplementary movie 10:} (cf.\ Fig.~\ref{fig7}c) Dynamics of an initially mixed suspension with active proportion $\alpha=1/2$ of pullers ($\beta=1$), at packing density $\phi=50\%$ and strong bottom-heaviness ($G_\text{bh}=100$). A lamellar ``sandwich''-like phase separation develops in which layers of passive spheres are pushed by layers of pullers, followed by a gap.

	\section*{Acknowledgments}
	T.I. was supported by the Japan Society for the Promotion of Science Grant-in-Aid for Scientific Research (JSPS KAKENHI Grants No. 21H04999 and No. 21H05308) and by JST CREST Grant No. JPMJCR25Q1.
	
	\section*{Declaration of Interests}
	The authors report no conflict of interest.
	
	\bibliographystyle{apsrev4-2}
	\bibliography{references}
	
\end{document}